\documentclass{article}

\usepackage{spconf,amsmath,graphicx,hyperref}

\usepackage{hyperref} 
\usepackage{comment}  
\usepackage{cite}
\usepackage{amsmath,amssymb,amsfonts}
\usepackage{algorithmic}
\usepackage{float}
\usepackage{xcolor,colortbl} % column color in table
\usepackage{amsmath,amsfonts} %
\usepackage{graphicx}%
\usepackage{textcomp}%
\usepackage{xcolor}%
\usepackage{url}
\usepackage{array}%
\usepackage{pifont}%

\usepackage{amsthm}%
\usepackage{bm}

\usepackage[linesnumbered,ruled,vlined]{algorithm2e}%
\usepackage{setspace}%

\usepackage{multirow}%
\usepackage{siunitx}
\usepackage{footnote}%
\usepackage[utf8]{inputenc}
\usepackage[english]{babel}

\usepackage{blindtext}

\usepackage{dblfloatfix}%

\usepackage{xcolor,colortbl} 
\usepackage{ctable} % for \specialrule command 
\graphicspath{ {image/} }

\usepackage{pifont}

\usepackage{booktabs}
\usepackage{makecell}
\usepackage[normalem]{ulem}
\usepackage{soul}

\DeclareUnicodeCharacter{2009}{\,}
\DeclareUnicodeCharacter{2248}{\ensuremath{\approx}}

\newcommand{\cmark}{\ding{51}}%
\newcommand{\xmark}{\ding{55}}%

\title{NLDSI-BWE: Non Linear Dynamical Systems-Inspired Multi Resolution Discriminators for Speech Bandwidth Extension}

\name{Tarikul Islam Tamiti \qquad Anomadarshi Barua}

\address{Department of Cyber Security Engineering, George Mason University, USA}

\begin{document}
\maketitle

\begin{abstract}

%Bandwidth Extension (BWE) reconstructs lost high frequencies due to bandwidth constraints for applications ranging from telecommunications to high-fidelity audio inference under limited resources. In this paper, we introduce nonlinear dynamical-systems-inspired discriminators, which model the inherent deterministic chaos of speech. Multi-Scale Recurrence Discriminator (MSRD) is designed to capture self-similarity dynamics, and Multi-Resolution Lyapunov Discriminator (MRLD) to capture nonlinear fluctuations and sensitivity to initial conditions. %Coupled with AP-BWE’s generator, MRAD, and MRPD, these discriminators deliver state-of-the-art (SoTA) results across objective and subjective evaluation benchmarks. 
%Through extensive design optimization and the use of depthwise separable convolutions in the discriminators, our framework surpasses prior SoTA while using only a fraction of the parameters. {\color{red}We believe this endeavor will serve as a stepping stone for future interdisciplinary research, where well-established theories from other disciplines synergize with AI.} %to solve problems in sound.}

%Bandwidth Extension (BWE) reconstructs lost high frequencies due to bandwidth constraints for applications ranging from telecommunications to high-fidelity audio inference under limited resources. 

In this paper, we design two nonlinear dynamical systems-inspired discriminators -- the Multi-Scale Recurrence Discriminator (MSRD) and the Multi-Resolution Lyapunov Discriminator (MRLD) -- to \textit{explicitly} model the inherent deterministic chaos of speech. MSRD is designed based on Recurrence representations to capture self-similarity dynamics. MRLD  is designed based on Lyapunov exponents to capture nonlinear fluctuations and sensitivity to initial conditions. 
Through extensive design optimization and the use of depthwise-separable convolutions in the discriminators, our framework surpasses prior AP-BWE model with a 44x reduction in the discriminator parameter count \textbf{($\sim$ 22M vs $\sim$ 0.48M)}.  To the best of our knowledge, for the first time, this paper demonstrates how BWE can be supervised by the subtle non-linear chaotic physics of voiced sound production to achieve a significant reduction in the discriminator size.

\end{abstract}


\vspace{-0.85em}
\section{Introduction}
\label{sec:intro}
\vspace{-0.5em}

\begin{figure*}[htbp]
  \centering
\includegraphics[width=0.85\textwidth,height=0.18\textheight]{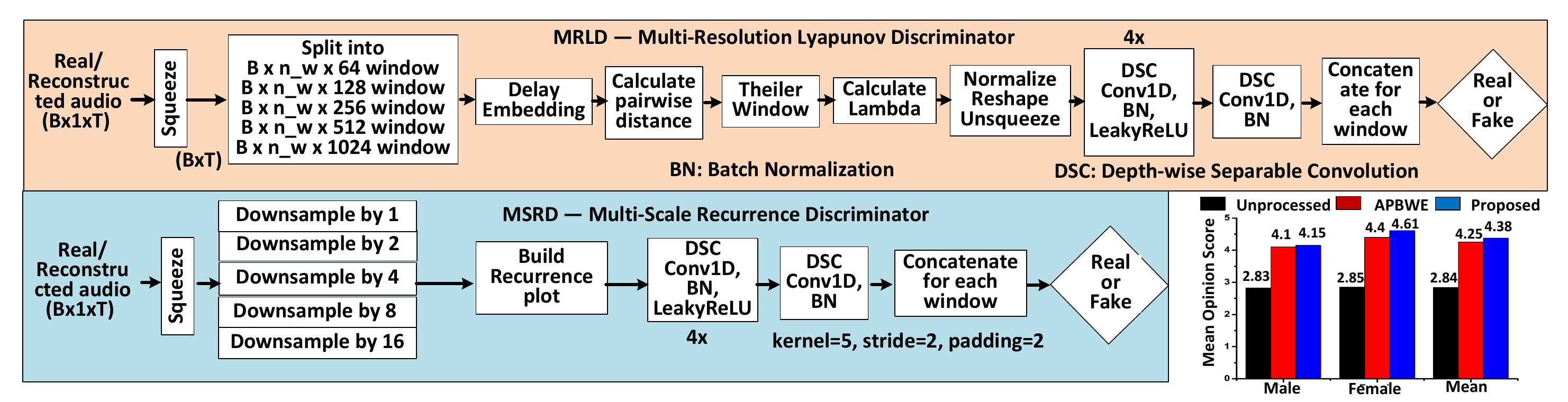}
\vspace{-01.550em}
  \caption{Our proposed NLDSI-BWE containing chaos-informed discriminators. (Right corner) Results of subjective MOS.}
  \label{fig:overall_architecture}
  \vspace{-01.760em}
\end{figure*}

Bandwidth Extension (BWE) aims to reconstruct high frequency content that is lost when speech is captured, stored, or transmitted at low sampling rates. %These missing HF contents significantly reduce the quality, faithfulness, and intelligibility of speech \cite{monson2014perceptual}. Thus, accurate BWE is vital for modern telephony and VoIP, hearing–assistive devices \cite{tamiti2025practical}, sound engineering, and audio enhancement \cite{tamiti2025high}, where the listening experience depends not only on spectral sharpness but also on natural prosody, intelligibility, and the absence of unnatural artifacts. 
While recent neural BWE frameworks \cite{kuleshov2017audio,ho2020denoising} provide improved spectral envelope, they often fail to reconstruct the micro-structure of voiced excitation and the rapid, nonlinear fluctuations characteristic of human speech. This failure results in over-smoothed spectra with reduced harmonic details that listeners perceive as dull, buzzy, or unstable \cite{jang2021univnet}. 

\textbf{Speech as a nonlinear dynamical system:} Vocal chord is a driven, self-sustained, viscous–elastic oscillator with aerodynamic coupling. Glottal flow and vocal-fold motion create quasi-periodic excitation around the fundamental frequency $f_0$, but with irregularities, turbulent components, and changes (e.g., breathy vs pressed voice) that are well modeled as \emph{deterministic chaos}  \cite{fitch2025applying,maccallum2009acoustic}. Two important chaotic features, such as (i) multi-scale recurrence of states in phase space (self-similarity at different time scales), and (ii) local divergent trajectories, can be estimated by Recurrence Plots \cite{eckmann1995recurrence} and Lyapunov Exponents \cite{oseledec1968multiplicative}. These features capture fine harmonic structure, micro-jitter, aperiodic bursts, and co-articulatory transitions, all of which contribute significantly to speech intelligibility and perceived naturalness. 

\textbf{What current Generative Adversarial Network (GAN) misses:} Most GAN-based BWE models rely on discriminators that only match distributional statistics of the waveform or its spectral magnitude (i.e., sometimes simple pitch periodicity) \cite{guo2022multi}, \cite{tamiti2025high}, \cite{lu2024multi}, \cite{nercessian2024dsp} but they rarely guide the generator to \textit{explicitly} reproduce the \emph{subtle non-linear chaotic features}, such as \emph{Recurence representations} and \emph{Lyapunov exponents}. 

AP-BWE \cite{lu2024towards} is a State-of-The-Art (SoTA) BWE model, which uses a parameter-heavy ($\sim$ 22 million (M) parameters) \emph{Multi-Period Discriminator (MPD)} \cite{kong2020hifi} to improve harmonic and periodic structures in the reconstructed speech.
In this paper, for the first time, we demonstrate that by explicitly replacing MPD with non-linear chaos-inspired discriminators, we can achieve \textit{better} performance with a 44x reduction in the discriminator parameter count \textbf{($\sim$ 22M vs $\sim$ 0.48M)}.

We name our proposed model NLDSI (Non Linear Dynamical Systems-Inspired) BWE. NLDSI-BWE demonstrates for the first time that if BWE is supervised by the subtle non-linear chaotic physics of voiced sound production, we can achieve better performance with a significant reduction in the discriminator size. We introduce two lightweight, nonlinear-dynamics-inspired discriminators: \textbf{\emph{MSRD}}, which maps the self-similarity structure by operating on multi-resolution recurrence representations; and \textbf{\emph{MRLD}}, which penalizes mismatches in local divergence rates by aligning Lyapunov Exponents across different resolutions. Both MSRD and MRLD are built from depthwise–separable convolutions with carefully chosen receptive fields and strides, while retaining long-context sensitivity through multi-scale processing to jointly capture coarse and fine grained chaotic dynamics \cite{shi2023time}. %This highly optimized design yields a drastic reduction in parameters while preserving and often improving performance.

\vspace{-0.95em}
\section{NLDSI-BWE Architecture}
\label{sec:Overall Architecture}
\vspace{-0.5em}
%Joint modeling of phase and amplitude are important for good perceptual quality \cite{lin2021two}. Hence, we design a dual-stream architecture in which amplitude and phase are simultaneously enhanced. The model takes as input a narrow‐band audio, which is upsampled  by sinc interpolation. Then \textbf{magnitude spectrogram} is generated by taking \textbf{logarithm of absolute value} of the results of Short Term Fourier Transform (STFT) and \textbf{phase spectrogram} is generated by taking the \textbf{angle} of the results of STFT. 

\vspace{-0.1em}
\subsection{Generator Architecture} 
\label{subsec:Generator Architecture}
\vspace{-0.5em}

To compare the capability of the proposed MRLD and MSRD, we keep the AP-BWE's generator \cite{lu2024towards} unchanged. The generator uses ConvNeXt \cite{liu2022convnet} as the core block with a criss-cross connection along with a dual stream for the exchange of amplitude and phase information. 

\vspace{-0.95em}
\subsection{Discriminator Architecture}
\label{subsec:discriminator Architecture}
\vspace{-0.5em}

\indent \textbf{a) Multi‐Resolution Lyapunov Discriminator (MRLD):} We introduce MRLD  (see Fig.~\ref{fig:overall_architecture} and Alg.~\ref{alg:MRLD}) based on Lyapunov Exponents (LE) \cite{oseledec1968multiplicative, wolf1985determining} to capture the rapid, nonlinear fluctuations and sensitivity to initial conditions in speech overlooked by SoTA. %The LE quantifies the average rate at which nearby trajectories separate. Therefore, MRLD penalizes mismatches in the Lyapunov spectra of real and generated signals and drives the generator to produce authentic deterministic chaotic behavior while producing more lifelike speech. 

\textbf{(Lines 1 to 7):} MRLD splits the waveform into multiple window lengths $w\in\{64,128,256,512,1024\}$ for multi-resolution coverage, creates delay embedding with dimension $m$ and delay $\tau$ to reconstruct the underlying state space, and searches for nearest neighbors after masking indices within a Theiler window $w_{\mathrm{th}}=m\tau$, which prevents trivial temporal self-matches and leakage. \textbf{(Lines 8 to 17):}  For each embedded point, we track the forward separation from its masked nearest neighbor for $k$ steps and average the log distances. Therefore,  a least-squares fit of this curve yields the local Lyapunov rate $\lambda$, which measures sensitivity to initial conditions. We stack the per chunk $\lambda$ to produce a compact 1-D exponent map in each $w$, which compresses dynamics without losing instability cues. We then feed it to a depthwise–separable 1-D CNN to learn patterns of dynamical cues from each resolution. MRLD concatenates per-resolution logits and feature maps for adversarial and feature matching losses to enforce that the generator matches Lyapunov statistics across different perspectives, and penalizes oversmoothed outputs, which preserves rapid, nonlinear speech dynamics.

\vspace{-1em}
\begin{algorithm}
\scriptsize
\caption{MRLD: (one input sample $x$)}
\label{alg:MRLD}
\begin{algorithmic}[1]
\REQUIRE Waveform $x\in\mathbb{R}^{T}$; window set $\mathcal{W}=\{64,128,256,512,1024\}$; embedding dimension $m$; delay $\tau$; small $\varepsilon>0$ for stability
\STATE \textbf{For each} $w\in\mathcal{W}$:
\STATE \quad Split $x$ into $n_w=\big\lfloor T/w\big\rfloor$ chunks $\{x^{(w)}_i\}_{i=1}^{n_w}$ of length $w$
\STATE \quad \textbf{For each} chunk $x^{(w)}_i$:
\STATE \qquad Set embedding length $M \leftarrow w-(m-1)\tau$
\STATE \qquad Form delay vectors $\mathbf{y}_j=\big[x_j, x_{j+\tau},\dots,x_{j+(m-1)\tau}\big]\in\mathbb{R}^{m}$ for $j=0,\dots,M-1$
\STATE \qquad Set Theiler window $w_{\mathrm{th}}\leftarrow m\tau$
\STATE \qquad Define allowed neighbor set $\mathcal{N}(j)=\{j'\,:\,|j-j'|>w_{\mathrm{th}}\}$
\STATE \qquad Find nearest neighbor index $\nu(j)=\arg\min_{j'\in\mathcal{N}(j)}\|\mathbf{y}_j-\mathbf{y}_{j'}\|_2$
\STATE \qquad Determine valid horizon $K \leftarrow M-\max_j \nu(j)-1$
\STATE \qquad \textbf{For} $k=0,\dots,K-1$:
\STATE \qquad\quad $d_k=\frac{1}{M-k}\sum_{j=0}^{M-k-1}\log\!\big(\,\|\mathbf{y}_{j+k}-\mathbf{y}_{\nu(j)+k}\|_2+\varepsilon\,\big)$
\STATE \qquad \textbf{end for}
\STATE \qquad Estimate Lyapunov rate $\displaystyle \lambda^{(w)}_i=\frac{\sum_{k=0}^{K-1} k\,d_k}{\sum_{k=0}^{K-1} k^2}$
\STATE \quad \textbf{end for}
\STATE \quad Aggregate results into exponent map $\lambda^{(w)}=\big[\lambda^{(w)}_1,\dots,\lambda^{(w)}_{n_w}\big]$
\STATE \quad Reshape $\lambda^{(w)}$ to $(1,1,n_w)$ and feed to the 1-D DSC SRLD
\STATE \quad Obtain logits $\ell^{(w)}$ and feature maps $F^{(w)}$
\STATE \textbf{end for}
\ENSURE Multi Resolution outputs $\{\ell^{(w)}\}_{w\in\mathcal{W}}$ and $\{F^{(w)}\}_{w\in\mathcal{W}}$ for \\ adversarial \& feature matching losses

\end{algorithmic}
\end{algorithm}
\vspace{-1em}

% --- MSRD (write-up + algorithm) ---

\begin{table*}[ht]
\centering
\scriptsize
\setlength{\tabcolsep}{1.8pt}
\renewcommand{\arraystretch}{0.65}
\begin{tabular}{l l
  ccc  ccc  ccc  ccc  ccc  ccc}
\toprule
\multirow{2}{*}{Method}
& \multirow{2}{*}{Size}
  & \multicolumn{3}{c}{NISQA-MOS}
  & \multicolumn{3}{c}{STOI}
  & \multicolumn{3}{c}{PESQ}
  & \multicolumn{3}{c}{SI-SDR}
  & \multicolumn{3}{c}{SI-SNR}
  & \multicolumn{3}{c}{LSD} \\
\cmidrule(lr){3-5}\cmidrule(lr){6-8}\cmidrule(lr){9-11}
\cmidrule(lr){12-14}\cmidrule(lr){15-17}\cmidrule(lr){18-20}
  &
  & 4–16 & 8–16 & 16–48
  & 4–16 & 8–16 & 16–48
  & 4–16 & 8–16 & 16–48
  & 4–16 & 8–16 & 16–48
  & 4–16 & 8–16 & 16–48
  & 4–16 & 8–16 & 16–48 \\
\midrule
Unprocessed   & -     & 2.79 & 3.67 & 4.43 & 0.55 & 0.61 & 0.61 & 1.15  & 1.51  & 1.41  & -11.03 & -8.07 & -6.07 & -10.53 & -7.62 & -5.63 & 3.27 & 2.27 & 2.85 \\
%EBEN (ICASSP,2023) \cite{hauret2023eben}   & 29.7M & 2.59 & 2.69 & 2.53     & 0.89 & 0.98 & 0.98      & 2.64  & 3.69  & 3.71      & 11.94  & 19.94 & 20.82      & 11.94  & 19.94 & 20.83      & 1.03  & 0.78  & 0.92     \\
EBEN (ICASSP,2023) \cite{hauret2023eben}   & 29.7M & 2.59 & 2.69 & 2.53     & 0.89 & 0.98 & 0.98      & 2.44  & 3.69  & 3.71      & 11.94  & 17.94 & 19.82      & 11.94  & 17.94 & 19.83      & 1.03  & 0.78  & 0.92     \\
%AERO (ICASSP,2023) \cite{mandel2023aero}  & 36.4M & 2.79 & 2.75 & 2.88 & 0.83 & 0.94 & 0.99  & 2.62  & 3.65  & 3.69  & 13.60  & 20.70 & 21.56 & 13.60  & 20.70 & 21.56 & 1.09  & 0.97  & 0.75 \\
AERO (ICASSP,2023) \cite{mandel2023aero}  & 36.4M & 2.79 & 2.75 & 2.88 & 0.83 & 0.94 & 0.99  & 2.42  & 3.65  & 3.69  & 12.60  & 17.70 & 19.56 & 12.60  & 17.70 & 21.56 & 1.09  & 0.97  & 0.75 \\

\rowcolor{red!30}
AP-BWE (TASLP, 2024)\cite{lu2024towards} & 72M   & 3.86 & 3.97 & 4.49 & 0.94 & 0.99 & 0.99  & 2.55  & 3.69  & 3.72  & 13.42  & 18.26 & 20.86 & 13.35  & 18.07 & 20.74 & 0.96  & 0.74  & 0.75 \\
\rowcolor{blue!30}
%NLDSI-BWE (proposed)          & 51.75M & 4.11 & 4.29 & 4.5 & 0.9417 & 0.998 & 0.991  & 2.34  & 3.69  & 3.64  & 12.76  & 17.59 & 19.49 & 12.67  & 17.43 & 19.44 & 0.99  & 0.7732  & 0.7864 \\
NLDSI-BWE (proposed)          & 51.75M & 4.11 & 4.29 & 4.5 & 0.94 & 0.99 & 0.99  & 2.54  & 3.69  & 3.70  & 12.76  & 17.59 & 19.49 & 12.67  & 17.43 & 19.44 & 0.97  & 0.73  & 0.77 \\

\bottomrule
\end{tabular}
\vspace{-01.3em}
\caption{Comparative analysis of baseline models over three extension ranges with our proposed NLDSI-BWE.}
\label{tab:Comparative_analysis}
\vspace{-02.5em}
\end{table*}

\textbf{b) Multi-Scale Recurrence Discriminator (MSRD):} We propose MSRD (see Fig.~\ref{fig:overall_architecture} and Alg.~\ref{alg:MSRD}), which leverages Recurrence Plots \cite{eckmann1987recurrence} to capture multi-scale temporal dependencies and hidden recurrent structures in speech. By modeling recurrence dynamics at multiple resolutions, MSRD highlights subtle periodicities and state transitions that are often missed by conventional SoTA approaches.

\textbf{(Lines 1 to 7):} MSRD evaluates the recurrence geometry (revisitation) of the waveform across multiple scales. For each scales $s \in \{1,2,4,8,16\}$, the waveform is downsampled by a stride $s$. To reduce computational complexity, if the decimated length exceeds a threshold of $L_{\max}=256$, we uniformly subsample to length $L_{\max}$. \textbf{(Lines 8 to 13):} We then generate a binarized recurrence plot (RP) using the pairwise absolute amplitude difference matrix $D$, where the mean of $D$ (diagonal included) acts as a threshold. In this way, a single-channel RP image is generated as an internal feature representation, which is then processed by a lightweight single-scale depthwise-separable 2-D CNN that generates patch logits and intermediate features. MSRD collects per-scale patch logits and features for adversarial and feature-matching objectives, giving the discriminator complementary access to the coarse-to-fine recurrence periodic structure.

\vspace{-1.3em}
\begin{algorithm}
\scriptsize
\caption{MSRD: (one input sample $x$)}
\label{alg:MSRD}
\begin{algorithmic}[1]
\REQUIRE Waveform $x\in\mathbb{R}^{T}$; scales $\mathcal{S}=\{1,2,4,8,16\}$; length cap $L_{\max}=256$.
\FOR{$s \in \mathcal{S}$}
    \STATE $x^{(s)} \leftarrow (x_0, x_s, x_{2s}, \ldots)$ \COMMENT{downsampled by stride $s$}
    \IF{$\lvert x^{(s)}\rvert > L_{\max}$}
        \STATE Uniformly subsample indices to obtain $\tilde{x}^{(s)} \in \mathbb{R}^{L_{\max}}$
    \ELSE
        \STATE $\tilde{x}^{(s)} \leftarrow x^{(s)}$
    \ENDIF
    \STATE $L \leftarrow \lvert \tilde{x}^{(s)} \rvert$
    \STATE $D_{p,q} \leftarrow \big\lvert \tilde{x}^{(s)}_{p} - \tilde{x}^{(s)}_{q} \big\rvert$ for $0 \le p,q < L$
    \STATE $\varepsilon^{(s)} \leftarrow \mathrm{mean}\big(\{D_{p,q}\}_{p,q}\big)$ \COMMENT{global mean threshold (with diagonal )}
    \STATE $\mathrm{RP}^{(s)}_{p,q} \leftarrow \mathbb{I}\!\big(D_{p,q} \le \varepsilon^{(s)}\big)$ \COMMENT{binary RP, single channel}
    \STATE $\ell^{(s)},\,F^{(s)} \leftarrow \mathrm{SSRD}\!\big(\mathrm{RP}^{(s)}\big)$ \COMMENT{2-D depthwise-separable CNN; patch \\logits + features}
\ENDFOR
\ENSURE MR outputs $\{\ell^{(s)}\}_{s\in\mathcal{S}}$ and $\{F^{(s)}\}_{s\in\mathcal{S}}$ for adversarial and \\ feature matching losses.
\end{algorithmic}
\end{algorithm}
\vspace{-1.5em}

\vspace{-01.0em} 
\subsection{Loss Functions}
\vspace{-0.52em}

\textbf{Generator loss functions:} Since we do not modify the AP-BWE generator, we use the same six different loss functions: magnitude loss  $\mathcal{L}_{\text{mag}}$, phase loss $\mathcal{L}_{\text{pha}}$, complex STFT loss $\mathcal{L}_{\text{com}}$, self-consistency loss $\mathcal{L}_{\text{stft}}$, feature matching loss $\mathcal{L}_{\text{fm}}$, and adversarial loss $\mathcal{L}_{\text{adv}}$. The total generator loss $\mathcal{L}_G$ is:

\vspace{-0.9em}
{\small
\begin{equation}
\mathcal{L}_G = \mathcal{L}_{\text{mag}} + \mathcal{L}_{\text{pha}} + 
\mathcal{L}_{\text{com}} + \mathcal{L}_{\text{stft}} + 
\mathcal{L}_{\text{fm}} + \mathcal{L}_{\text{adv}}.
\end{equation}
}
\vspace{-1.26em}

\vspace{-1.50em}
{
\small
\begin{equation}
\mathcal{L}_D = \sum_r \mathcal{L}_D^{\text{MRLD}}  + 
\sum_s \mathcal{L}_D^{\text{MSRD}}  + 
\sum_r \mathcal{L}_D^{\text{MRAD}}   + 
\sum_r \mathcal{L}_D^{\text{MRPD}}   
\label{eqn:discriminitorloss}
\end{equation}
}
\vspace{-0.8em}

\textbf{Discriminator loss functions:} %Each discriminator \(D_d\) is trained using a hinge loss objective, encouraging to differentiate between generated and ground truth audioss. This loss specializes discriminators to become powerful critics of any unnatural patterns by matches with the perceptual distribution of real speech. 
Each discriminator \(D_d\) is trained using a hinge loss objective. The total discriminator loss $\mathcal{L}_D$ is shown in Eqn. \ref{eqn:discriminitorloss}, where $\mathcal{L}_D^{\text{MRLD}}$, $\mathcal{L}_D^{\text{MSRD}}$, $\mathcal{L}_D^{\text{MRAD}}$, and $\mathcal{L}_D^{\text{MRPD}}$ are MRLD, MSRD, Multi-Resolution Amplitude Discriminator (MRAD), and Multi-Resolution Phase Discriminator (MRPD) losses, respectively, for each resolution/scale. We replace only MPD by MRLD and MSRD, and reuse MRAD and MRPD losses from AP-BWE \cite{lu2024towards}. %The training alternates between minimizing \(\mathcal{L}_D\)  and \(\mathcal{L}_G\) using AdamW optimizers and exponential learning rate schedulers.

\vspace{-0.95em}
\subsection{Evaluation Metrics}
\vspace{-0.5em}

To comprehensively evaluate the proposed NLDSI-BWE in terms of intelligibility, fidelity, and perceived quality, we use Log-Spectral Distance (LSD), Short-Time Objective Intelligibility (STOI), Perceptual Evaluation of Speech Quality (PESQ), Scale-Invariant Signal-to-Distortion Ratio (SI-SDR), and Non-Intrusive Speech Quality Assessment - Mean Opinion Score (NISQA-MOS) \cite{mittag2021nisqa}. 

%To comprehensively evaluate the reconstructed audio, we use five metrics: Log Spectral Distance (LSD) for spectral distortion, Short-Time Objective Intelligibility (STOI)  for intelligibility, Perceptual Evaluation of Speech Quality (PESQ)  for perceived quality, Scale-Invariant Signal-to-Distortion Ratio (SI-SDR)  for overall signal-noise distortion, and Non-Intrusive Speech Quality Assessment - Mean Opinion Score (NISQA-MOS) to estimate the perceived quality.

\vspace{-0.95em}
\subsection{Dataset, Preprocessing, and Hyperparameter} 
\label{subsec:training_setup}
\vspace{-0.52em}

We use the VCTK Corpus (v0.92)~\cite{yamagishi2019cstr}, which contains 110 multi-accent English speakers with 400 utterances each at 16/48 kHz. We load 16/48 kHz files, convert to mono channel, remove silence parts, downsample to simulate band-limited audio, sinc interpolate, align length-wise, and cache audio. We train with a batch size of 16 using the AdamW optimizer (\(\beta_{1}=0.8\), \(\beta_{2}=0.99\)) and a weight decay of \(0.01\). The learning rate is initialized at \(2\times10^{-4}\) and decays exponentially at each epoch with a factor of \(0.999\). Models are trained for 50 epochs, with each epoch taking approximately 25 minutes. Experiments are conducted on four NVIDIA RTX~4090 GPUs and Intel Xeon Silver~4310 CPUs. %We create an Anaconda virtual environment with  Python~3.9.21, PyTorch~2.0.0{+}cu118, Torchaudio~0.15.0{+}cu118, Torchvision~0.15.0{+}cu118, and the CUDA Toolkit~11.8.0 to run codes.

\vspace{-0.99em}
\subsection{Comparative Analysis with Baselines}
\label{subsec:comparative-baselines}
\vspace{-0.45em}

Across the three frequency ranges (4–16, 8–16, 16–48\,kHz), our proposed NLDSI-BWE consistently delivers the \textbf{best speech quality (NISQA-MOS)} while maintaining \emph{almost SoTA intelligibility (STOI)} and LSD (see Table \ref{tab:Comparative_analysis}). Specifically, it provides the highest NISQA-MOS in all three frequency ranges, outperforming EBEN, AERO, and AP-BWE. STOI is also higher for NLDSI-BWE than EBEN/AERO and on par with AP-BWE (0.94/0.99/0.99). In terms of spectral fidelity, NLDSI-BWE performs similarly to AP-BWE on PESQ and LSD and retains a slight edge on SI-SDR and SI-SNR.  (e.g., 20.86 vs 19.49 and 20.74 vs 19.44 for 16-48 kHz). EBEN and AERO exhibit slightly higher SI-SDR/SI-SNR at highband fills, but \textbf{lag substantially in NISQA-MOS and LSD}, suggesting that their signal fidelity does not fully translate to human-perceived quality. Therefore, AP-BWE is the best-performing model among baselines, and we compare NLDSI-BWE with the best-performing AP-BWE. 

Please note that NLDSI-BWE achieves the highest NISQA -MOS and STOI, and similar LSD and PESQ with only 51.75M parameters, which is 28\% lower than AP-BWE’s 72M parameter count. Given the higher NISQA-MOS at comparable STOI, PESQ, and LSD, this indicates a favorable accuracy–capacity balance. The parameter reduction happens mainly for replacing MPD by our newly designed chaos-inspired discriminators -- MRLD and MSRD. This proves our important point that explicitly integrating non-linear chaotic physics into discriminators can give better performance with a smaller model size (see Sections \ref{subsec:Discriminator_Ablation} and \ref{subsec:Comparison-with-MPD} for details).

\vspace{-0.95em}
\subsection{Discriminator Ablation: Key Observations}
\label{subsec:Discriminator_Ablation}
\vspace{-0.52em}

We provide an ablation study of discriminators in Table \ref{table:discriminator_ablation_study}.

\noindent\textbf{a) Row \textcircled{\scriptsize 1}:}
Without any discriminators, the generator-only model (U-Net) gives NISQA-MOS=3.33, STOI=0.88, LSD = 1.256, and SI-SNR=9.25. This is a baseline with limited intelligibility and noticeable spectral error. 

\noindent\textbf{b) Rows \textcircled{\scriptsize 2}-\textcircled{\scriptsize 4}:} MRLD-only (Row\textcircled{\scriptsize 2}) and MSRD-only (Row\textcircled{\scriptsize 3}) models slightly raise STOI and keep SI-SNR close to baseline, but drop NISQA-MOS and do not consistently improve LSD. Combining them (Row\textcircled{\scriptsize 4}) does not recover NISQA-MOS and further worsens LSD. \textbf{Hypothesis:} MRLD /MSRD, when used alone, pressures the generator toward dynamical/recurrence plausibility but lacks amplitude/phase cues. This hurts perceived quality despite marginal intelligibility gains. 

\noindent\textbf{c) Rows \textcircled{\scriptsize 5}-\textcircled{\scriptsize 6}:} Adding amplitude/phase critics, such as  MRAD+MRPD (Row\textcircled{\scriptsize 5}) from AP-BWE sharply improves perceptual quality (NISQA-MOS=4.07) and reduces spectral error (LSD=1.126), though STOI and SI-SNR fall. This makes MRAD+MRPD as the must-have critic for achieving good results across evaluation metrics.  
Introducing MSRD (Row\textcircled{\scriptsize 6}) further improves LSD and NISQA-MOS while modestly recovering SI-SNR. \textbf{Hypothesis:} MRAD/MRPD provides strong magnitude/phase cues, boosting NISQA-MOS. MSRD then adds long-horizon structure regularization that stabilizes spectra and mitigates over-smoothing. 

\noindent\textbf{d) Rows \textcircled{\scriptsize 7}-\textcircled{\scriptsize 8}} Changing MRPD for MRLD alongside MRAD (Row\textcircled{\scriptsize 7}) yields the best non-MPD spectral LSD=1.10 with improved STOI (0.87 vs\ Rows\textcircled{\scriptsize 5}--\textcircled{\scriptsize 6}) with slightly lower NISQA-MOS. The full quartet MRAD + MRPD +   MRLD+MSRD (Row\textcircled{\scriptsize 8}) reaches a strong overall balance. %\textit{N\!-MOS}=4.13, \textit{LSD}=1.11, \textit{STOI}=0.867, and \textit{SI\!-SNR}=7.63. 
\textbf{Hypothesis:} MRLD fine-tune MRAD's oversmoothed spectra by enforcing chaotic details, while MRPD+MSRD counterbalance each other by removing noisy phase and revisiting to previous states. \textbf{Row\textcircled{\scriptsize 8} gives our proposed well-balanced discriminator combination for NLDSI-BWE.}  %yielding a well-rounded critic ensemble with robust perceptual and spectral outcomes.%(phase fidelity vs.\ multi-scale recurrence), yielding a well-rounded critic ensemble with robust perceptual and spectral outcomes.

\vspace{-01.195em}
\subsection{Comparison with MPD}
\label{subsec:Comparison-with-MPD}
\vspace{-0.42em}

\noindent \textbf{a) Rows \textcircled{\scriptsize 9}-\textcircled{\scriptsize 10}:} MPD+MRLD (Row\textcircled{\scriptsize 9}) underperforms perceptually (NISQA-MOS=3.48) relative to our non-MPD quartet, shown in Row \textcircled{\scriptsize 8}, due to the absence of amplitude and phase cues. MPD+MRAD+MRPD (Row\textcircled{\scriptsize 10}) achieves the best NISQA-MOS (4.11) within Rows \textcircled{\scriptsize 9}-\textcircled{\scriptsize 10}  but with weaker intelligibility/fidelity (STOI=0.8537, SI-SNR=6.67) and moderate spectral error (LSD=1.11). \textbf{Hypothesis:}  Parameter -heavy MPD can model perceptual sharpness (higher NISQA-MOS) but does not enforce micro-dynamical or multi-scale recurrence cues as explicitly as MRLD/MSRD. 

\noindent\textbf{b) Row \textcircled{\scriptsize 8} vs Rows \textcircled{\scriptsize 9}-\textcircled{\scriptsize 10}:} Compared to AP-BWE model, having MPD+MRAD+MRPD (Row\textcircled{\scriptsize 10}), our NLDSI-BWE, having MRAD+MRPD+MRLD+MSRD (Row\textcircled{\scriptsize 8}), achieves higher performance for all five metrics.

\vspace{-0.65em}
\begin{table}[ht!]
  %\centering
  \scriptsize
  \setlength{\tabcolsep}{1.3pt}
\begin{tabular}{l c c c c c c c c c c}
\toprule
SL & MPD & MRAD & MRPD & MRLD & MSRD
   & LSD & STOI & PESQ & SNR & N-MOS \\[-0.5ex]
\hline
\multicolumn{11}{c}{}\\[-6.5pt]
\multicolumn{11}{c}{\textbf{Baselines and single additions}}\\[1pt]
\hline
1 & \xmark & \xmark & \xmark & \xmark & \xmark
  &  1.2557& 0.8799   & 1.8450  & 9.2548 & 3.3261 \\
2 & \xmark & \xmark & \xmark & \cmark & \xmark
  & 1.2467 & 0.8814 & 1.8725 & 9.1822  & 2.3973 \\
3 & \xmark & \xmark & \xmark & \xmark & \cmark
  & 1.2618 & 0.8832 & 1.9142 & 9.2069 & 2.3600 \\
4 & \xmark & \xmark & \xmark & \cmark & \cmark
  &1.2709  & 0.8825 & 1.8557 & 9.2236 & 2.3710 \\
\hline
\multicolumn{11}{c}{}\\[-6.5pt]
\multicolumn{11}{c}{\textbf{MRAD/MRPD pair (w/ and w/o MSRD)}}\\[1pt]
\hline
5 & \xmark & \cmark & \cmark & \xmark & \xmark
  &1.1261   & 0.8663 & 1.5945   & 7.6817  &  4.0728\\
6 & \xmark & \cmark & \cmark & \xmark & \cmark
  & 1.1221 & 0.8631 &  1.5939 & 8.238 & 4.1935 \\
\hline
\multicolumn{11}{c}{}\\[-6.5pt]
\multicolumn{11}{c}{\textbf{Trios (MRAD+MRPD+MRLD), $\pm$ MSRD}}\\[1pt]
\hline
7 & \xmark & \cmark & \xmark & \cmark & \xmark
  & 1.1058 & 0.8697 & 1.6643 &8.2749 & 3.9645 \\
\rowcolor{blue!30}
8 & \xmark & \cmark & \cmark & \cmark & \cmark
  & 1.1112 & 0.8669 &  1.6146& 7.6332 & 4.1312\\
\hline
\multicolumn{11}{c}{}\\[-6.5pt]
\multicolumn{11}{c}{\textbf{MPD comparisons}}\\[1pt]
\hline
9  & \cmark & \xmark & \xmark & \cmark & \xmark
   & 1.1975  & 0.8648   &  1.64  & 7.574  &  3.48\\
\rowcolor{red!30}
10 & \cmark & \cmark & \cmark & \xmark & \xmark
   & 1.1101 & 0.8537 & 1.56   & 6.671  & 4.11 \\

\hline
\multicolumn{11}{c}{}\\[-6.5pt]
\multicolumn{11}{c}{\textbf{Parameter comparison (per discriminator, not cumulative)}}\\[2pt]
\hline
11 & {\color{magenta}22M} & 600.2k & 600.2k & {\color{blue}235.5k} & {\color{blue}247.7k}
   &  & & & & \\
\bottomrule
\end{tabular}
\vspace{-0.9765em}
\caption{Ablation study on discriminators for 2$\rightarrow$16 kHz. Here, N-MOS = NISQA-MOS and SNR = SI-SNR.}
%Row\,\textcircled{\scriptsize 8} (MRAD+MRPD+MRLD+MSRD) achieves a strong composite of \textit{NISQA-MOS}/\textit{STOI}/\textit{LSD}/\textit{SNR} with far fewer parameters than MPD-heavy setups. 
%Replacing MPD (22M) with MRLD+MSRD ($\approx$483.2k) inside an otherwise identical stack reduces parameters by $\sim$45$\times$ (pairwise) or $\sim$13$\times$ when comparing full stacks (Row\,\textcircled{\scriptsize 10} vs.\ Row\,\textcircled{\scriptsize 8}).}
\label{table:discriminator_ablation_study}
\vspace{-1.2em}
\end{table}

\noindent\textbf{c) Row \textcircled{\scriptsize 11} (Parameter efficiency):} An MPD uses $\sim$22M parameters, whereas our designed MRLD+MSRD together uses a total of $\sim$483.2k parameters ($235.5$k+$247.7$k), which indicates 44x parameter reduction. The size of the full quartet (MRAD+MRPD+MRLD+MSRD) in our proposed NLDSI-BWE  is $\sim$1.684M parameters, which is $\sim$13.77x smaller than AP-BWE's trios (MPD+MRAD+MRPD, $\sim$23.2M).

\textbf{Row \textcircled{\scriptsize 8}}, \textbf{Row \textcircled{\scriptsize 10}}, and \textbf{Row \textcircled{\scriptsize 11}} indicate that adding chaotic micro-dynamics (MRLD) and multi-scale recurrence structure (MSRD) to amplitude/phase critics can match MPD’s perceptual gains while offering better intelligibility and signal fidelity with only a fraction of parameters. \textit{This observation will encourage the community to adopt our model in resource-constrained edge devices for the BWE task.}

\noindent\textbf{Overall takeaway:} Amplitude/phase critics drive perceptual gains; dynamical/recurrence critics (MRLD/MSRD) improve temporal structure, reduce oversmoothing, and produce crispier pleasant sounds. The quartet in Row\textcircled{\scriptsize 8} delivers the best composite across all five metrics without the MPD’s parameter and computational burden.

\vspace{-0.8em}
\begin{table}[ht]
\centering
\scriptsize
\setlength{\tabcolsep}{3pt}
\renewcommand{\arraystretch}{0.85}

\begin{tabular}{lcccccc}
\toprule
Freq range &  LSD & STOI  & PESQ   & SI-SDR & SI-SNR & NISQA-MOS \\
\midrule
2-16 & 1.11 & 0.8669 & 1.6146 & 7.63 & 7.62 & 4.13\\
2-48 & 1.1281 & 0.83185 & 1.194 & 7.5966 & 7.5923 & 4.0123\\
4-16 & 0.9904 & 0.9417 & 2.3415 & 12.76 & 12.677 & 4.1069\\
8-16 & 0.7732 & 0.998 & 3.6894 & 17.59 & 17.433 & 4.2948 \\
8-48 & 0.9355 & 0.9963 & 2.4228 & 15.5252 & 15.424 & 4.5171\\
12-48 & 0.8498 & 0.998 & 3.2083 & 18.0015 & 17.9058 & 4.51056 \\
16-48 & 0.7864 & 0.9981 & 3.6443 & 19.4958 & 19.44 A& 4.5023\\
24-48 & 0.653 & 0.9987 & 4.1583 & 22.6708 & 22.65603 & 4.51637\\

\bottomrule
\end{tabular}
\vspace{-01.2em}
\caption{Performance over different frequency ranges.}
\label{table:freqrange}
\end{table}

\vspace{-1.3em}
\subsection{Results Across Different Bands} 
\vspace{-0.5em}

Table \ref{table:freqrange} indicates a clear pattern: \emph{as the gap between the narrow and target band reduces, performance improves}. Therefore, the broadest reconstruction (2-48\,kHz) is the hardest, giving the poorest scores, and the narrowest reconstruction (24-48\,kHz) is the easiest, giving the best scores. Table \ref{table:freqrange}  also indicates that all five metrics improve between 2-48\,kHz and 24-48\,kHz when the gap between the narrow and target bands is reduced.

\vspace{-01.73em}
%\subsection{Computational Complexity and Real-Time Performance}
\subsection{Computational Complexity}
\label{subsec:Computational Complexity}
\vspace{-0.52em}

Computational complexity and real-time performance is shown in Table \ref{tab:computation} by using Generator Parameters in Millions (GP), Discriminator Parameters in Millions (DP), Multiply Accumulate Operations (MACs), Floating Point Operations per second (FLOPs), and Real-Time Factor (RTF) across two different frequency ranges. As only the generator is used during inference, the MACs, FLOPs, and RTF are the same as the AP-BWE baseline. The GP is the same for both AP-BWE and NLDSI-BWE, as we do not change the generator design, while the DP is reduced significantly due to the replacement of parameter-heavy MPD with MRLD+MSRD.

\vspace{-0.4em}
\begin{table}[ht]
\centering
\scriptsize
\setlength{\tabcolsep}{2.1pt}
\renewcommand{\arraystretch}{0.85}
\begin{tabular}{lcccccc}
\toprule
Model & Fq. Range & GP & DP & MACs (M) & FLOPs (M) & RTF (GPU)  \\
\midrule

\rowcolor{red!30}
AP-BWE & 4-16 kHz & 29.76 &42.3 & 14236.65 & 28473.31  & 0.0023x \\
\rowcolor{red!30}
AP-BWE & 16-48 kHz & 29.76&42.3 &14236.65 & 28473.31 &  0.0025x \\
\rowcolor{blue!30}
NLDSI-BWE & 4-16 kHz & 29.76& 1.68&14236.65 & 28473.31  & 0.0023x \\
\rowcolor{blue!30}
NLDSI-BWE & 16-48 kHz &29.76& 1.68&14236.65 & 28473.31 &  0.0025x \\

\bottomrule
\end{tabular}
\vspace{-0.74em}
\caption{Computational complexity of NLDSI-BWE. The hardware configuration is provided in Section \ref{subsec:training_setup}.}
\label{tab:computation}
\vspace{-01.4em}
\end{table}

\vspace{-01.44em}
\subsection{Subjective Analysis}
\label{subsec:Subjective Test}
\vspace{-0.52em}

Subjective comparison of NLDSI-BWE against AP-BWE and unprocessed audio is conducted by a selected panel of 10 persons. We use 5-point (1=bad to 5=excellent) Mean Opinion Score (MOS) ratings for the subjective evaluation. In the bottom-right of Fig. \ref{fig:overall_architecture}, we show the bar-chart of MOS results separately for male and female speakers and their mean. NLDSI-BWE significantly outperforms AP-BWE for female speakers and overall. However, the performance gain for male speakers is very negligible. From this result, we can comment that the proposed NLDSI-BWE may reconstruct high-frequency contents effectively, as typically female voices contain higher frequencies than their male counterparts. Similarly, results provide strong evidence that our proposed NLDSI-BWE consistently generates perceptually higher audio, which is favored by a wide range of listeners.

%AP-BWE performs better for only male speakers, while . In the Pairwise preference test, NLDSI-BWE outperforms SOTA AP-BWE by a margin of 11\%. The detailed explanation of subjective evaluation is presented in Appendix \ref{subsec:subj_eval_details}. 

\begin{comment}
 We expose a new speech threat that adversaries can recover intelligible audio up to 8 kHz from severely aliased pressure sensor data, having a sampling frequency greater than 500 Hz. Using our HVAC-EAR, an attacker can secretly listen to natural conversation behind the wall that is the least unexpected. Moreover, we comprehensively evaluate HVAC-EAR using five metrics that have not been done before. However, HVAC-EAR is tested on only English dataset, works up to 1.2 m distance and does not perform well if the sampling frequency is less than 500 Hz.
\end{comment}

\vspace{-01.3em}
\section{Conclusion and Limitations}
\label{sec:Conclusion}
\vspace{-0.52em}

We propose NLDSI-BWE, which is a complex-valued, dual-stream model and has non-linear systems-inspired discriminators. We propose MRLD (chaotic divergence) and MSRD (recurrence geometry) to enforce perceptually natural- sounding and phase-consistent reconstructed audios while reducing oversmoothing phenomena with a reduced set of parameters.  However, we only test the model with the VCTK dataset rather than multiple datasets in noise-free settings. In multi-lingual and cross-speaker settings, the generalization ability is not tested. We will handle these in our upcoming work. Moreover, due to the introduction of non-linear complicated calculations in discriminators, the training time is slightly higher compared to AP-BWE.%(i.e., 15 minutes/epoch vs 20 minutes/epoch).  

\vspace{-01.3em}
\section{Compliance with Ethical Standards}
\label{sec:Compliance}
\vspace{-0.52em}

This research study was conducted retrospectively using human subject data made available in open access by The University of Edinburgh’s Data Share repository
 \cite{sarfjoo2018}. Ethical approval was not required, as confirmed by the license attached to the open-access data.

\begin{comment}
{\color{red}We conducted a listening study to assess perceived quality across three extension ranges (2–16 kHz, 12–48 kHz, and 24–48 kHz) using both 5-point Mean Opinion Score (MOS) ratings and pairwise “peer” preference tests. A panel of 12 trained listeners evaluated each condition—Unprocessed, AP-BWE, our Proposed method, and the Clean reference—presented in randomized order. For MOS, listeners rated each sample on a scale from 1 (bad) to 5 (excellent); for the peer test, they selected their preferred sample in head-to-head comparisons.}
\end{comment}
%As shown in Table X, the Unprocessed signals received the lowest MOS across all ranges (mean scores around 1.8–2.1), reflecting severe high-frequency loss. AP-BWE yielded moderate gains (MOS ≈ 2.9–3.2), but remained noticeably below the Clean reference (MOS ≈ 4.6–4.8). Our Proposed method achieved substantial improvements—mean MOS of approximately 3.8–4.2—narrowing the gap to the Clean condition by over 50 \% at all bands.

%In the peer tests, the Proposed model was preferred over AP-BWE in more than 75 \% of pairwise trials across every frequency range, and even occasionally over the Clean reference when minor artifacts in the original recording were less audible. These results demonstrate that by explicitly modeling nonlinear dynamical cues, our approach not only restores high-frequency energy but also delivers perceptually more natural and preferred audio.

\bibliographystyle{IEEEtran}
\bibliography{version_3}
\end{document}